\documentclass[12pt]{article}
\newlength{\dinwidth}
\newlength{\dinmargin}
\usepackage{graphicx}          

\setkeys{Gin}{keepaspectratio} 

\begin{document}

\def\ap#1#2#3   {{\em Ann. Phys. (NY)} {\bf#1} (#2) #3.}   
\def\apj#1#2#3  {{\em Astrophys. J.} {\bf#1} (#2) #3.} 
\def\apjl#1#2#3 {{\em Astrophys. J. Lett.} {\bf#1} (#2) #3.}
\def\app#1#2#3  {{\em Acta. Phys. Pol.} {\bf#1} (#2) #3.}
\def\ar#1#2#3   {{\em Ann. Rev. Nucl. Part. Sci.} {\bf#1} (#2) #3.}
\def\cpc#1#2#3  {{\em Computer Phys. Comm.} {\bf#1} (#2) #3.}
\def\err#1#2#3  {{\it Erratum} {\bf#1} (#2) #3.}
\def\ib#1#2#3   {{\it ibid.} {\bf#1} (#2) #3.}
\def\jmp#1#2#3  {{\em J. Math. Phys.} {\bf#1} (#2) #3.}
\def\ijmp#1#2#3 {{\em Int. J. Mod. Phys.} {\bf#1} (#2) #3}
\def\jetp#1#2#3 {{\em JETP Lett.} {\bf#1} (#2) #3}
\def\jpg#1#2#3  {{\em J. Phys. G.} {\bf#1} (#2) #3.}
\def\mpl#1#2#3  {{\em Mod. Phys. Lett.} {\bf#1} (#2) #3.}
\def\nat#1#2#3  {{\em Nature (London)} {\bf#1} (#2) #3.}
\def\nc#1#2#3   {{\em Nuovo Cim.} {\bf#1} (#2) #3.}
\def\nim#1#2#3  {{\em Nucl. Instr. Meth.} {\bf#1} (#2) #3.}
\def\np#1#2#3   {{\em Nucl. Phys.} {\bf#1} (#2) #3}
\def\pcps#1#2#3 {{\em Proc. Cam. Phil. Soc.} {\bf#1} (#2) #3.}
\def\pl#1#2#3   {{\em Phys. Lett.} {\bf#1} (#2) #3}
\def\prep#1#2#3 {{\em Phys. Rep.} {\bf#1} (#2) #3}
\def\prev#1#2#3 {{\em Phys. Rev.} {\bf#1} (#2) #3}
\def\prl#1#2#3  {{\em Phys. Rev. Lett.} {\bf#1} (#2) #3}
\def\prs#1#2#3  {{\em Proc. Roy. Soc.} {\bf#1} (#2) #3.}
\def\ptp#1#2#3  {{\em Prog. Th. Phys.} {\bf#1} (#2) #3.}
\def\ps#1#2#3   {{\em Physica Scripta} {\bf#1} (#2) #3.}
\def\rmp#1#2#3  {{\em Rev. Mod. Phys.} {\bf#1} (#2) #3}
\def\rpp#1#2#3  {{\em Rep. Prog. Phys.} {\bf#1} (#2) #3.}
\def\sjnp#1#2#3 {{\em Sov. J. Nucl. Phys.} {\bf#1} (#2) #3}
\def\shep#1#2#3 {{\em Surveys in High Energy Phys.} {\bf#1} (#2) #3}
\def\spj#1#2#3  {{\em Sov. Phys. JEPT} {\bf#1} (#2) #3}
\def\spu#1#2#3  {{\em Sov. Phys.-Usp.} {\bf#1} (#2) #3.}
\def\zp#1#2#3   {{\em Zeit. Phys.} {\bf#1} (#2) #3}

\title{\vspace{5cm}
\bf{ Evidence for no shrinkage in
elastic photoproduction of $J/\Psi$
}
\vspace{2cm}}

\author{
 {\bf Aharon Levy} \\ 
{\small \sl School of Physics and Astronomy}\\ {\small \sl Raymond and 
Beverly Sackler Faculty of Exact Sciences}\\
  {\small \sl Tel Aviv University, Tel Aviv, Israel}
}
\date{ }
\maketitle

\begin{abstract}
  The differential cross section $d\sigma/dt$ of elastic
  photoproduction of $J/\Psi$ at different $\gamma p$ center of mass
  energies $W$ are used to determine the properties of the exchanged
  trajectory in the reaction $\gamma p \to J/\Psi p$. In the region
  $|t| <$ 1 GeV$^2$, the resulting trajectory is $\alpha(t)$ = (1.153
  $\pm$ 0.027) + (-0.001 $\pm$ 0.072)$t$. The vanishing value of the
  slope of the trajectory is an experimental evidence for no shrinkage
  and thus for the perturbative nature of the exchange.
\end{abstract}

\vspace{-18cm}
\begin{flushleft}
TAUP 2468--97 \\
December 1997 \\
\end{flushleft}

\setcounter{page}{0}
\thispagestyle{empty}
\newpage  

%

%
%
%
%
%
%
%
\section{Introduction}

High energy interactions are a combination of soft and hard
interactions. It is however not completely clear how to quantify these
regions. Hard interactions are usually thought of as those which can
be completely calculated by perturbative QCD. The soft ones are in the
nonperturbative domain and usually need some
experimental/phenomenological input to carry out calculations. In most
cases one is in a situation where there is interplay between soft and
hard processes~\cite{afs}.

A particular case where this interplay is present are diffractive
processes and one discusses soft and hard diffraction. The soft
diffraction is characterized as a process in which one exchanges a
Pomeron trajectory with the properties of having an intercept
$\alpha_0$ of about 1.08 and a slope $\alpha^\prime$ of 0.25
GeV$^{-2}$. This trajectory is also sometimes referred to as the 'soft
Pomeron' or the 'nonperturbative Pomeron'.

The hard diffractive processes seen e.g. in deep inelastic scattering
reactions, seem to be described by a trajectory with a larger
intercept than that of the 'soft Pomeron'. One calls such a trajectory
'hard Pomeron'. Although clearly there is just one Pomeron and it is
the one invented by Gribov to describe the rise of the total
hadron--hadron cross section, we will nevertheless use the notions of
'soft Pomeron' and 'hard Pomeron' for the sake of clarity.  The
expectations of the 'hard Pomeron' trajectory are to have a larger
intercept than that of the 'soft Pomeron' and a smaller slope. The
reason for the smaller slope is connected to Gribov
diffusion~\cite{gribov-dif} and will be elaborated on later.

A very clean diffractive reaction in which only a Pomeron can be exchanged
is the elastic photoproduction of $J/\Psi$
\begin{equation}
\gamma p \to J/\Psi p.
\end{equation}
The cross section for this reaction rises faster than expected from a
'soft' diffractive reaction. While the cross section of elastic
photoproduction of the three lightest vector mesons $\rho^0$, $\omega$
and $\phi$ rises like $W^{0.22}$\cite{HERA-lightvm}, that of the
$J/\Psi$ rises like $W^{0.64\pm0.13}$~\cite{h1} or
$W^{0.92\pm0.14}$~\cite{zeus}. This indicates that the intercept of
the exchanged 'object' in the latter reaction is larger than 1.08 but
in order to measure this intercept from the cross section behaviour
one needs information also on the slope $\alpha^\prime$.

How can one get information on $\alpha^\prime$? One way is by looking
for the shrinkage of the diffractive peak with increasing energy. This
is done by assuming a functional form of
the differential cross section $d\sigma/dt$, for example a
single exponential $\exp(bt)$, and then studying the dependence of the
exponential slope $b$ on $W$,
\begin{equation}
b = b_0 + 4\alpha^\prime \ln W.
\end{equation}

A more direct way is by measuring the whole trajectory directly. This
is done by studying the $W$ dependence of $d\sigma/dt$ at fixed $t$
values,
\begin{equation}
\frac{d\sigma}{dt} = f(t) (W^2)^{[2\alpha(t)-2]},
\label{eq:dsdt}
\end{equation}
where $f(t)$ is a function of $t$ only.  Thus one can determine
$\alpha(t)$ at each $t$ value and by fitting a linear form to the
trajectory,
\begin{equation}
\alpha(t) = \alpha_0 + \alpha^\prime t,
\label{eq:trajectory}
\end{equation}
one finally determines the intercept and the slope of the trajectory.

In order to use such a direct determination of the Pomeron trajectory,
one needs to study a reaction which is driven only by a Pomeron
exchange. The best examples of such reactions are either $\gamma p \to
\phi p$ or $\gamma p \to J/\Psi p$. Since the $\phi$ and the $J/\Psi$
are pure $s\bar{s}$ and $c\bar{c}$ states, respectively, the reactions
can only proceed via a Pomeron exchange, even at low $W$ values, as
the exchange of secondary trajectories is suppressed in these cases.

In the present note we report on the results of a study of the energy
dependence of the differential cross section for elastic
photoproduction of $J/\Psi$ in order to measure $\alpha(t)$
using equation~(\ref{eq:dsdt}). Using these values and
equation~(\ref{eq:trajectory}) we want to check in particular whether
the trajectory has a slope similar to that of the non--perturbative
Pomeron leading to shrinkage or whether in the case of the $J/\Psi$
vector meson, the heavy mass of its constituents leads to a
perturbative type of process in which shrinkage is suppressed.

\section{The data}

The data used in the present analysis come from four different
experiments. The lowest energy data used here are from EMC~\cite{emc}
at an average energy of $W$ = 16.2 GeV. The data are presented as
$d\sigma/dp_T^2$ for a cut of $z \geq$ 0.95, where $z$ is the fraction
of the photon carried by the $J/\Psi$, corrected for coherence at low
$p_T^2$. A sum of two exponentials were fit to the data yielding
$d\sigma/dp_T^2=(89\pm 15)\exp[(-5.2\pm 0.6)p_T^2] + (2.3\pm
0.9)\exp[(-0.66\pm 0.14)p_T^2]$. This expression was used for $p_T^2 <$
1 GeV$^2$ since in this region the approximation $|t| \approx p_T^2$
is a good one. Also for $p_T^2 <$ 1 GeV$^2$ the contamination of
events where the nucleon dissociates is small in the elastic sample.

The second data set is taken from the E401 experiment~\cite{e401}
where $d\sigma/dt$ has been measured at $W$ = 16.8 GeV for the elastic
process in the region $|t| <$ 1 GeV$^2$. This experiment was performed
with real photons enabling a direct measurement of $t$. A careful
study was performed to isolate the elastic events. An exponential 
in $t$ and $t^2$ was fit to the data resulting in $d\sigma/dt=(80\pm
13)\exp[(5.6\pm 1.2)t+(2.9\pm 1.3)t^2)]$.

The H1 collaboration~\cite{h1} measured $d\sigma/dp_T^2$ for the
elastic channel at $W$ = 85 GeV and for $p_T^2 <$ 1 GeV$^2$ the data
can be represented by a single exponential $\exp(-bp_T^2)$ with a
slope of $b=(4.0\pm 0.2\pm 0.2)$ GeV$^{-2}$. The effects of using
$p_T^2$ instead of $|t|$ are small and accounted for in the systematic
error on the slope.

The highest energy point used in this analysis, $W$ = 90 GeV, is from
the ZEUS experiment~\cite{zeus} which measured $d\sigma/dt$ for the
elastic channel. Also these data can be described for $|t| <$ 1
GeV$^2$ by a single exponential in $t$, yielding a slope of $b=4.6\pm
0.4^{+0.4}_{-0.6}$ GeV$^{-2}$.

\section{Results}

Figure 1 shows the differential cross section for the elastic process
$\gamma p \to J/\Psi p$, as function of the $\gamma p$ center of mass
energy $W$, for five fixed $t$ values, $|t|$=0.1, 0.3, 0.5, 0.7 and 0.9
GeV$^2$. For each of the five $t$ values a fit of the form,
\begin{equation}
\frac{d\sigma}{dt} = f(t) W^{[4\alpha(t) - 4]},
\end{equation}
was performed and the value of $\alpha(t)$ was determined. These
values are plotted in figure 2 as function of $t$ and seem to be
independent of $t$. A linear fit of the form $\alpha(t) = \alpha_0 +
\alpha^\prime t$ yields
\begin{eqnarray}
\alpha_0 &=& 1.153 \pm 0.027, \\
\alpha^\prime &=& -0.001 \pm 0.072 \ \ {\rm GeV}^{-2}.
\end{eqnarray}
For comparison, the trajectory of the Donnachie--Landshoff 'soft'
Pomeron~\cite{dl} is also shown in the figure as a dashed line.

If one assumes that $f(t)$ is due to the electromagnetic form factor
of the proton, one can fit all 20 data points to the form
\begin{equation}
\frac{d\sigma}{dt} = \frac{C}{\left(1 + \frac{t}{m^2} \right)^4}
W^{(4\alpha_0 + 4\alpha^\prime t - 4)}.
\end{equation}
A very god fit to the data is obtained with the following results
\begin{eqnarray}
\alpha_0 &=& 1.161 \pm 0.023, \\
\alpha^\prime &=& -0.029 \pm 0.054 \ \ {\rm GeV}^{-2}.
\end{eqnarray}
The value of the scale parameter from the fit is $m^2$ = 0.72 $\pm$ 0.23
GeV$^2$, consistent with the accepted value of 0.71 GeV$^2$ used to
describe the electromagnetic form factor. The value of the parameter
$C$ is 14 $\pm$ 6, with units such that $d\sigma/dt$ is given in
nb/GeV$^2$.

\section{Discussion and conclusions}

The resulting value of $\alpha^\prime \approx$ 0 is evidence to the
fact that there is no shrinkage of $d\sigma/dt$ in the process $\gamma
p \to J/\Psi p$. Shrinkage is a characteristic behaviour of
diffractive processes dominated by 'soft' Pomeron exchange. The value
of $\alpha^\prime$ is connected to the inverse value of the average
$k_T$ of the partons involved in the exchange. A process can start at
the photon in a small configuration with a high $k_T$ value but as a
result of Gribov diffusion~\cite{gribov-dif} the partons interacting
with those of the proton become 'soft' due to their random diffusion
and thus the value of $\alpha^\prime$ is non zero, leading to
shrinkage. The fact that no shrinkage is observed in elastic
photoproduction of $J/\Psi$ indicates that Gribov diffusion is
unimportant in this process at the presently available $W$ values, and
the average $k_T$ remains large. Such a behaviour is
expected~\cite{cfs} from processes governed by the 'perturbative
Pomeron' whose values of $\alpha(t)$ are expected to remain larger
than 1. This makes the process a 'hard' one, fully calculable in
perturbative QCD.

In conclusion, the trajectory of the 'object' exchanged in the process
$\gamma p \to J/\Psi p$ was determined to be $\alpha(t) = (1.153 \pm
0.027) + (-0.001 \pm 0.072) t$, giving evidence for no shrinkage in
the process. This can be interpreted as indication that for the
reaction $\gamma p \to J/\Psi p$ diffusion from small to large size
configuration is only a small correction up to $W$ = 90 GeV. Thus the
process is 'hard' and fully calculable in perturbative QCD.

\section{Acknowledgment}

It is a pleasure to acknowledge fruitful discussions with Halina
Abramowicz, Lonya Frankfurt and Mark Strikman. Thanks are due also to
colleagues from the HERA experiments, Beate Naroska from H1, Song
Ming Wang and Riccardo Brugnera from ZEUS for helpful discussions of
the data.

This work was partially supported by the German--Israel Foundation
(GIF), by the U.S.--Israel Binational Foundation (BSF) and by the
Israel Science Foundation (ISF).

\begin{figure}[h]
\begin{center}
  \includegraphics [bb=17 103 539 729,width=\hsize,totalheight=18cm]
  {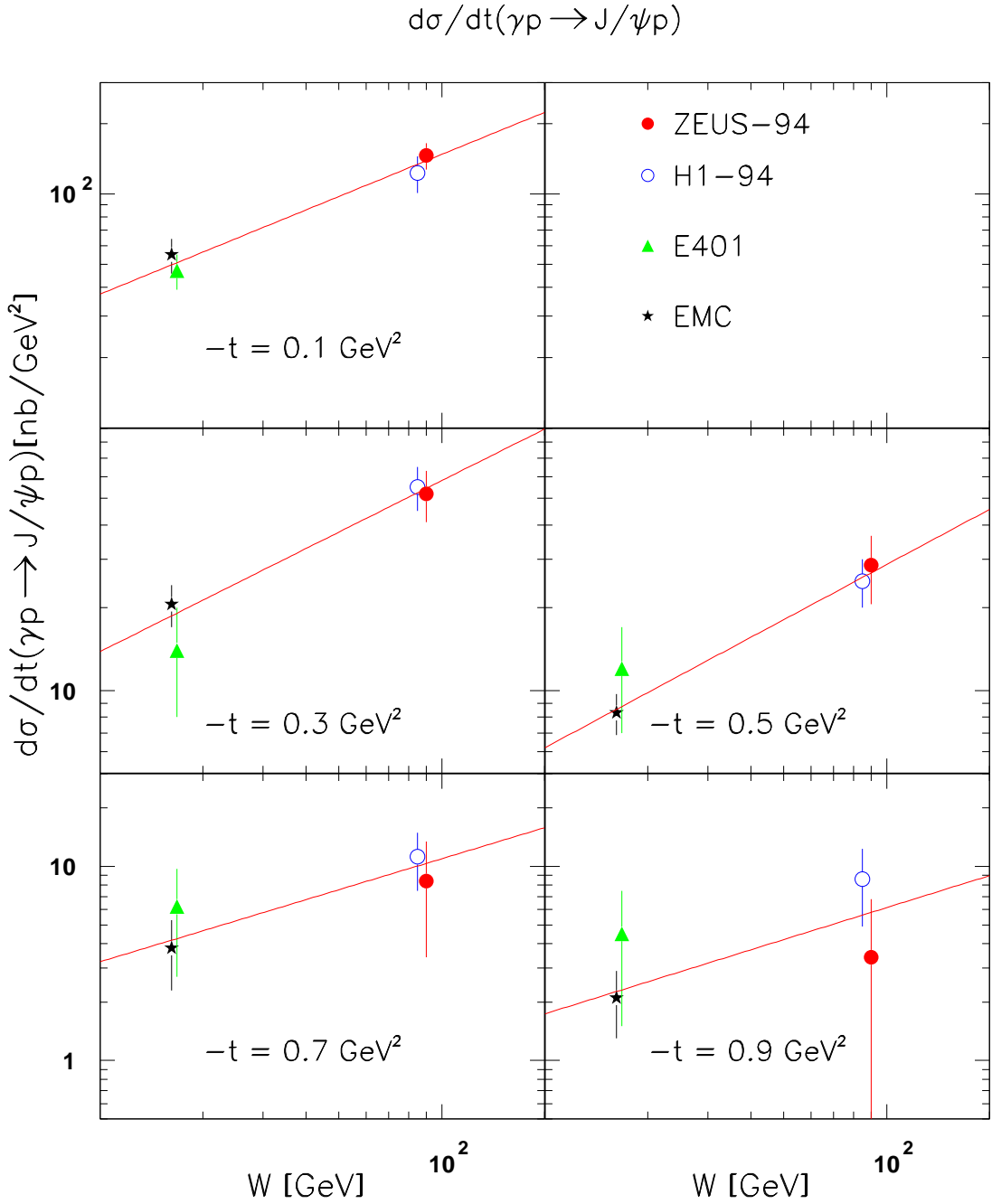}
\end{center}
\vspace{-.5cm}
\caption
{
The differential cross section $d\sigma/dt$ of the reaction $\gamma p
  \to J/\Psi p$ as function of $W$ for fixed values of $t$, as
  indicated in the figures. The line is the result of a fit of the
  form $d\sigma/dt=A(t)W^{[4\alpha(t)-4]}$ to the data.
}
\label{fig:dsdt}
\end{figure}

\begin{figure}[h]
\begin{center}
  \includegraphics [bb=20 69 536 727,width=\hsize,totalheight=18cm]
  {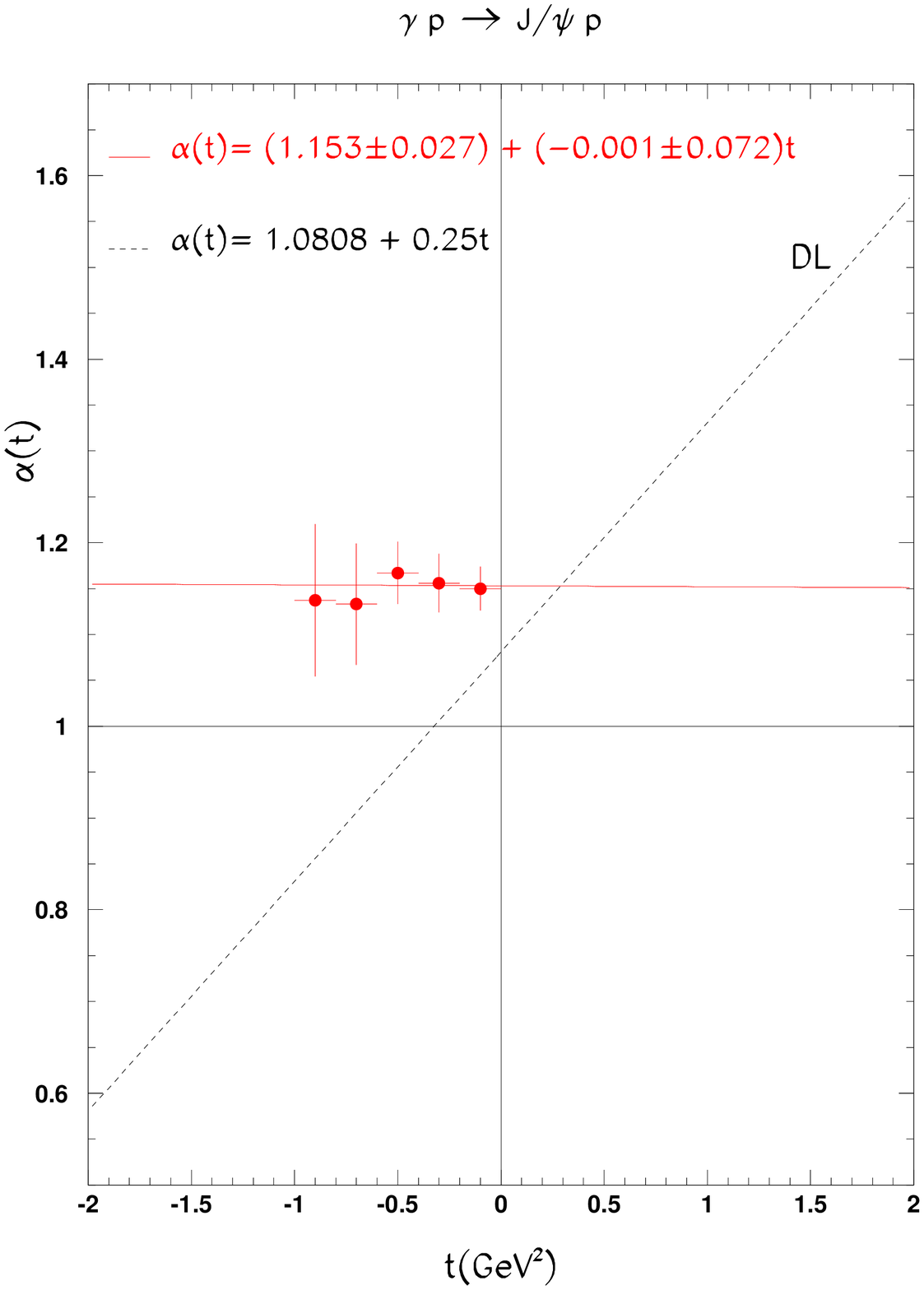}
\end{center}
\vspace{-.5cm}
\caption
{ 
The values $\alpha(t)$ of the exchanged trajectory in the reaction
$\gamma p \to J/\Psi p$ as function of $t$. The solid line is the
result of a fit of the form $\alpha(t)=\alpha_0+\alpha^\prime t$ to
the data. The dashed line is the trajectory of the
Donnachie--Landshoff (DL) Pomeron.
}
\label{fig:alphaj}
\end{figure}

\end{document}